\documentclass[aps,pre, twocolumn, floatfix,superscriptaddress,showpacs,showkeys]{revtex4-1}
\usepackage[dvipsnames]{xcolor}
\usepackage{epsf,amsmath,amssymb,verbatim,color,multirow,pifont,graphicx,cleveref,tabularx}
\crefname{equation}{Eq.}{Eqs.}
\crefname{section}{Sec.}{Secs.}
\crefname{figure}{Fig.}{Figs.}
\crefname{table}{Table}{Tables.}

\newcommand{\ro}{\rho_0}
\newcommand{\ps}{P_{S}}
\newcommand{\pw}{P_{W}}
\newcommand{\pr}{P_{IR}}

\newcommand{\dsqrt}{\big\langle d_0^{-1/2} \big\rangle_+}

\begin{document}

\title{Two golden times in two-step contagion models}
\author{Wonjun Choi}
\affiliation{CCSS, CTP and Department of Physics and Astronomy, Seoul National University, Seoul 08826, Korea}
\author{Deokjae Lee}
\affiliation{CCSS, CTP and Department of Physics and Astronomy, Seoul National University, Seoul 08826, Korea}
\author{J. Kert\'esz}
\affiliation{Center for Network Science, Central European University, Budapest, Hungary}
\affiliation{Department of Theoretical Physics, Budapest University of Technology and Economics, Budapest, Hungary}
\author{B. Kahng}
\email{bkahng@snu.ac.kr}
\affiliation{CCSS, CTP and Department of Physics and Astronomy, Seoul National University, Seoul 08826, Korea}
\date{\today}

\begin{abstract}
The two-step contagion model is a simple toy model for understanding pandemic outbreaks that occur in the real world. The model takes into account that a susceptible person either gets immediately infected or weakened when getting into contact with an infectious one. As the number of weakened people increases, they eventually can become infected in a short time period and a pandemic outbreak occurs. The time required to reach such a pandemic outbreak allows for intervention and is often called golden time. Understanding the size-dependence of the golden time is useful for controlling pandemic outbreak. Here we find that there exist two types of golden times in the two-step contagion model, which scale as $O(N^{1/3})$ and $O(N^{\zeta})$ with the system size $N$ on Erd\H{o}s-R\'enyi networks, where the measured $\zeta$ is slightly larger than $1/4$. They are distinguished by the initial number of infected nodes, $o(N)$ and $O(N)$, respectively. While the exponent $1/3$ of the $N$-dependence of the golden time is universal 
even in other models showing discontinuous transitions induced by cascading dynamics, the measured $\zeta$ exponents are all close to $1/4$ but show model-dependence. It remains 
open whether or not $\zeta$ reduces to $1/4$ in the asymptotically large-$N$ limit. 
\end{abstract}

%\pacs{89.75.Hc, 64.60.ah, 05.10.-a}
\maketitle

\section{Introduction}

Epidemic spread of diseases and rumors and their control and containment have become a central issue in recent years as the real world becomes ``smaller.'' It is a general observation that there is a slow phase in the spreading process before the sudden pandemic outbreak~\cite{review_epidemics}. This slow period is called golden time as it allows for intervention, which is much more difficult after the disease becomes global. Modeling of epidemic spread with essential factors is necessary to control catastrophic outbreaks within this golden time. To this end, several epidemic models have been investigated on complex networks, for instance, the susceptible--infected--removed (SIR) model~\cite{sir,sir_newman} and the susceptible--infected--susceptible (SIS) model~\cite{sis}. Analytical and numerical studies of those models revealed that a continuous phase transition occurs on Erd\H{o}s-R\'enyi (ER) random networks~\cite{ER}. Thus, abrupt pandemic outbreaks on a macroscopic scale, which often occur in the real world, cannot be reproduced using those models. 

Considerable effort has been devoted recently to construct mathematical models that exhibit a discontinuous epidemic transition at a finite transition point on complex networks. A natural way is to appropriately extend the conventional SIR and SIS models. For instance, an extended SIR model includes more than one infected state of different pathogens that are cooperatively activated in contagion: A person who is suffering from the flu can be more easily infected by pneumonia. This model is referred to as a cooperative contagion model~\cite{grassberger_nphy}. Similar instances include a two-step contagion process. A patient becomes weakened first and then becomes sick. This model is referred to as the susceptible--weakened--infected--removed (SWIR) model~\cite{janssen,janssen_spinoal,grassberger_pre,hasegawa1,hasegawa3,chung,choi_2016,choi_multi}. 
In another instance of modified SIR models, a network evolves by rewiring links at a certain rate during the spread of contagion~\cite{sir_rewire}. The rewiring takes into account the mobility of humans. Then, epidemic spread can be accelerated as the rewiring rate is increased, which can lead to a discontinuous transition representing the pandemic outbreak. 

When diseases spread, we need to keep susceptible people separate from infected patients or vaccinate the susceptible people before the diseases spread on a macroscopic level. A recent study~\cite{universal} showed that for the SWIR model on ER networks, a system exhibits a long latent period (called a golden time) within which measures can be taken, beyond which the disease spreads explosively over the system at a macroscopic level. Estimating the golden time is important for the prevention of pandemic outbreaks. Moreover, it is necessary to get early-warning signals if a critical threshold is  approached~\cite{indicator}.

It was revealed~\cite{choi_2016,universal} that when a disease starts spreading from a single node, the golden time $n_c$ scales as $n_c(N)\sim N^{\zeta}$ with $\zeta=1/3$ at the epidemic threshold. 
Here we reconsider this problem and represent the pattern of disease transmission using a nonlinear mapping. We show that the linear and nonlinear terms of the nonlinear mapping separately behave dynamically well. The linear term is responsible for one-step contagion without weakened states and the nonlinear term describes the two step contagion, which includes weakened state. Thus, the previous result of $N^{1/3}$ for the golden time is consistent with the characteristic size of the giant cluster generacolorted in the SIR model~\cite{bennaim}, thus it has got verified within this new framework.  
Next, we consider another case, which is the main concern of this paper, in which an epidemic starts to spread from endemic multiple seeds of $O(N)$ on ER networks also at the epidemic threshold. In this case, long latent period appears not immediately but after some characteristic time. Thus fluctuations induced by the stochastic process of disease transmission in the early time heavily affect the behavior during the latent period, which changes the measured exponent $\zeta$ to a value slightly larger than 1/4. We estimate this scaling behavior using the saddle-node bifurcation theory~\cite{book} and discuss the underlying mechanism. 

Similar size dependences of mean cascading time at a transition point were studied for other cascade dynamics models such as $k$-core percolation~\cite{kcore1,kcore2,kcore3,kcore4, kcore5,kcore_goltsev} and cascading failure model on interdependent network (CFoIN)~\cite{buldyrev,zhou,son,baxter_mcc,bianconi}.  It was found~\cite{buldyrev,zhou} that in the CFoIN, the mean cascading time is proportional to $N^{1/3}$ or $N^{1/4}$ depending on the way of choosing the transition points. Refs.~\cite{kcore4, kcore5} showed that the exponent 1/3 is also obtained in $k$-core percolation. Thus, the scaling behavior of $N^{1/3}$ is robust. However, for the  $\zeta > 1/3$ case, a different scaling behavior with $\zeta \approx 0.280$~\cite{grassberger_mcc} was 
numerically obtained for a surface growth model effectively equivalent to the CFoIN. 

Here we extend our formalism of nonlinear mapping used in the SWIR model to other models such as $k$-core percolation and the threshold model~\cite{watts,dodds}. We show that when the cascade starts from a fixed number of multiple seeds $O(N)$, the golden times for both models also become proportional to $N^{\zeta}$, where $\zeta$ are estimated to be slightly larger than 1/4 within our simulation range and those values are different to each other, suggesting non-universal behavior. However, we cannot exclude the possibility $\zeta=1/4$ in large-$N$ limit. We shall discuss this point in Sec. IV.

This paper is organized as follows: We first introduce the SWIR model and set up the evolution equation of the epidemic dynamics in Sec. II. Next, we derive a nonlinear mapping for the epidemic spread from a single seed in Sec. IIIA. We show that the roles of the linear and nonlinear terms are well separated. In Sec. IIIB, we derive a similar nonlinear mapping for the multiple-seed case, and show how the multiplicative feature of the fluctuations of epidemic spreading affects scaling of the golden time. In Sec. IV, we obtain the golden times of the multiple-seed case for $k$-core percolation and the threshold model and show that the numerical values of $\zeta$ are slightly larger than $1/4$. We also discuss the possibility of $\zeta=1/4$ in the thermodynamic limit. In Sec. V, we discuss the origin of the puzzle in view of nonlinear dynamics theory. A summary is presented in Sec. VI. 

\section{The SWIR model}

The SWIR model is a generalization of the SIR model by including two sates, a weakened state (denoted as $W$) and an infected state ($I$), between the susceptible state ($S$) and recovered state ($R$), instead of a single infected state $I$ alone, as in the SIR model. Nodes in state $W$ are involved in the reactions $S+I \to W+I$ and $W+I\to 2I$, which occur in addition to the reactions $S+I\to 2I$ and $I\to R$ in the SIR model. At each discrete time step $n$, the following processes are performed. (i) All the nodes in state $I$ are listed in random order. (ii) The states of the neighbors of each node in the list are updated sequentially as follows: If a neighbor is in state $S$, it changes its state in one of two ways: either to $I$ with probability $\kappa$ or to $W$ with probability $\mu$. If a neighbor is in the state $W$, it changes to $I$ with probability $\eta$, where $\kappa$, $\mu$, and $\eta$ are the contagion probabilities for the respective reactions. (iii) All nodes in the list change their states to $R$. This completes a single time step, and we repeat the above processes until the system reaches an absorbing state in which no infectious node is left in the system. The reactions are summarized as follows:
\begin{eqnarray}
\rm{S+I}  &\buildrel{\kappa}\over \longrightarrow& {\rm I+I}, \label{si_ii} \\
\rm{S+I} &\buildrel{\mu}\over \longrightarrow& \rm{W+I}, \label{si_wi} \\
\rm{W+I}  &\buildrel{\eta}\over \longrightarrow& \rm{I+I}, \label{wi_ii} \\
\rm{I}  &\buildrel{1}\over \longrightarrow& \rm{R}. \label{i_r} 	
\end{eqnarray}

In an absorbing state, each node is in one of three states, the susceptible,  weakened, or recovered state. We define $\ps(\ell)$ as  the conditional probability that a node remains in state $S$ in the absorbing state,
provided that it has $\ell$ neighbors in state $R$ and was originally in state S.
This means that the node remains in state $S$ even though it has been in contact $\ell$ times with these $\ell$ neighbors in state $I$ before they change their states to $R$. Thus, we obtain
\begin{equation}
\ps(\ell) = (1-\kappa-\mu)^{\ell}.
\label{p_s}
\end{equation}
Next, $P_{W}(\ell)$ is similarly defined as the conditional probability that a randomly selected susceptible node is in state $W$ after it contacts $\ell$ neighbors in state $I$ before they change their states to $R$. The probability $\pw(\ell)$ is given as 
\begin{equation}
\pw(\ell) = \sum_{n=0}^{\ell-1} (1-\kappa-\mu)^{n} \mu (1-\eta)^{\ell-n-1}.
\label{p_w}
\end{equation}
%where $\nu$ is the probability of W becoming I by single contact with neighboring I.
Finally, $\pr(\ell)$ is the conditional probability that a node has been infected in any state, either $I$ or $R$,  provided that it was originally in state $S$ and its $\ell$ neighbors are in state $R$ in the absorbing state.
Using the relation $\ps(\ell)+\pw(\ell)+\pr(\ell)=1$, one can determine $\pr(\ell)$ in terms of $\ps$ and $\pw$.  

On a network with a degree distribution $P_d$, we consider the case in which the initial densities of susceptible, weakened, and infectious nodes are given as $s_0$, $w_0,$ and $i_0$, provided that $s_0 +w_0+i_0=1$. The order parameter $m$, the density of nodes in state $R$ after the system falls into an absorbing state, is given using the local tree approximation as
\begin{equation}
m=i_0+\sum_{q=1}^{\infty}P_{d}(q)\Big(s_0 f_q(u)+w_0 g_q(u)\Big),
\label{eq:m_s}
\end{equation}
where
\begin{equation}\label{key}
f_q(u)=\sum_{\ell=1}^{q} \binom{q}{\ell} u^{\ell}(1-u)^{q-\ell} \pr(\ell)
\end{equation}
\begin{eqnarray}\label{key}
g_q(u)&=&\sum_{\ell=1}^{q} \binom{q}{\ell} u^{\ell}(1-u)^{q-\ell}\Big(1-(1-\eta)^{\ell}\Big) \nonumber \\
&=&1-(1-\eta u)^q 
\end{eqnarray}
and $u$ is the probability that an arbitrarily chosen edge leads to a node in state $R$ or $I$ but not infected through the chosen edge in the absorbing state. We define $u_n$ similarly to $u$ but at time step $n$. The probability $u_{n+1}$ can be derived from $u_n$ as follows:
\begin{equation} \label{eq:sce}
u_{n+1}=i_0+\sum_{q=1}^{\infty}\dfrac{qP_d(q)}{z}\Big(s_0 f_{q-1}(u_{n})+w_0 g_{q-1}(u_n)\Big),
\end{equation}
where $z\equiv\sum_{q} qP_d(q)$ is the mean degree of the network and the factor $qP_d(q)/z$ is the probability that a node connected to a randomly chosen edge has degree $q$. As $n\to\infty$, $u_n$ converges to $u$.
\section{Golden times in the SWIR model}

\subsection{The single-seed case}

\begin{figure}[h]
	\centering
	\includegraphics[width=0.9\linewidth]{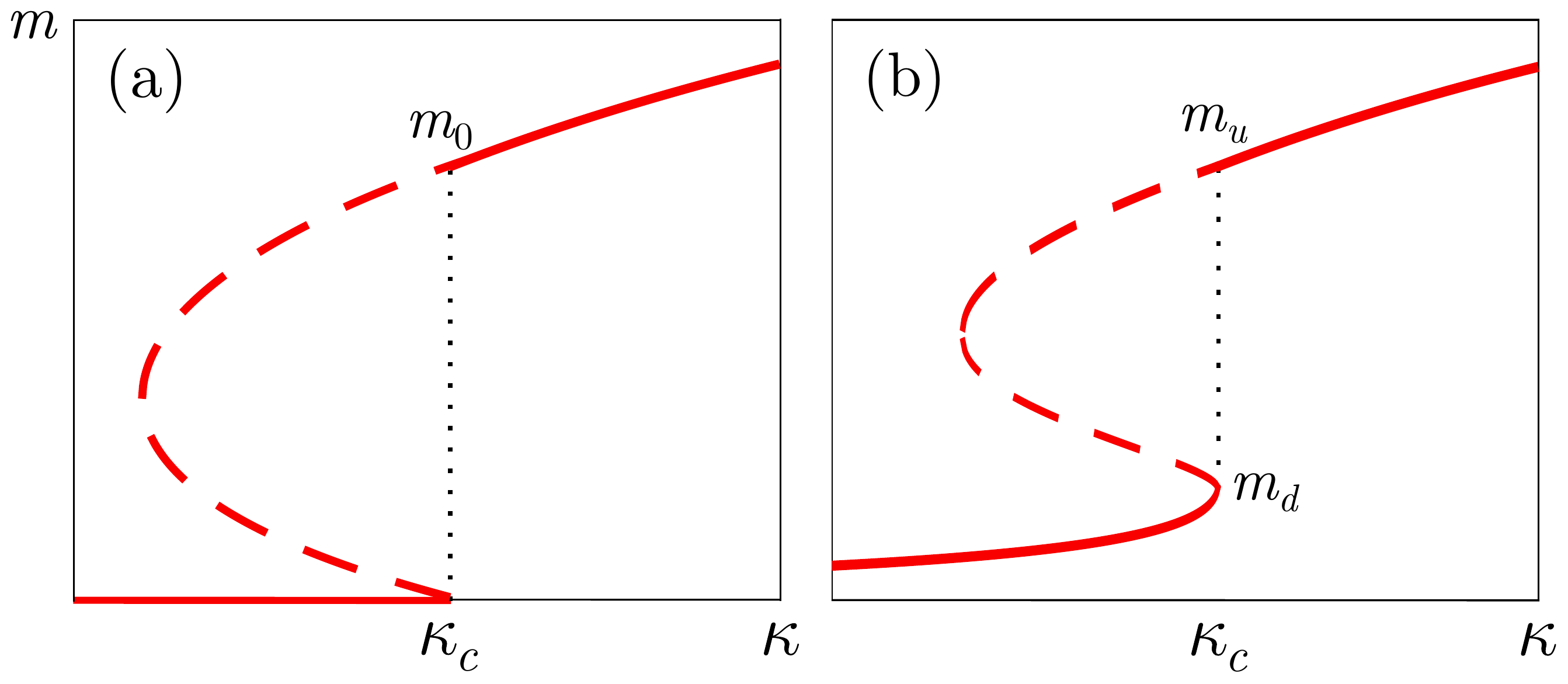}
	\caption{Schematic plots of the order parameter $m(\kappa)$ versus $\kappa$ for (a) $i_0=0$ (the single-seed case) and (b) $i_0 > 0$ (the multiple-seed case). (a) Even though $\kappa$ is increased, the order parameter remains zero up to a transition point $\kappa_c$. However, at $\kappa_c$, it remains at $m=0$ or jumps to $m_0$. For (b), as $\kappa$ is increased, $m(\kappa)$ gradually increases from $i_0$ to a finite value $m_d$ at $\kappa_c$. At $\kappa_c$, $m(\kappa)$ remains at $m_d$ or jumps to $m_u$.}
	\label{fig:op}
\end{figure}

First, we consider the case in which the initial number of infectious nodes is $o(N)$; that is, $i_0=w_0=0$ and $s_0=1$ in the thermodynamic limit. In this case, the SWIR model exhibits a mixed-order transition~\cite{choi_2016} at a transition point $\kappa_c$ when the mean degree is larger than a critical value. The order parameter displays a discontinuous transition from $m(\kappa_c)=0$ to $m_0$, whereas other physical quantities such as the outbreak size distribution exhibit a critical behavior. The behavior of the order parameter $m(\kappa)$ as a function of $\kappa$ is schematically shown in Fig.~\ref{fig:op}(a).

We are interested in how infected nodes spread as a function of the cascade step $n$ when the order parameter jumps.  
As a particular case, when the network is an ER network having a degree distribution that follows the Poisson distribution, i.e., $P_d(q)=(q+1)P_d(q+1)/z=z^q e^{-z}/q!$, where $z$ is the mean degree, Eq.~(\ref{eq:sce}) is reduced as follows:
\begin{eqnarray} \label{eq:f(r_n)}
u_{n+1}&=&1-\Big(1-\dfrac{\mu}{\kappa + \mu - \eta}\Big)e^{-(\kappa + \mu)z u_n} - \dfrac{\mu}{\kappa +\mu -\eta}e^{-\eta z u_n}. \nonumber \\
&\equiv& F(u_n)
\end{eqnarray}
We remark that on ER networks, $u_n$ in the limit $n\to \infty$ becomes equivalent to $m$ obtained from Eq.~(\ref{eq:m_s}).
\begin{figure*}
	\centering
	\includegraphics[width=0.9\linewidth]{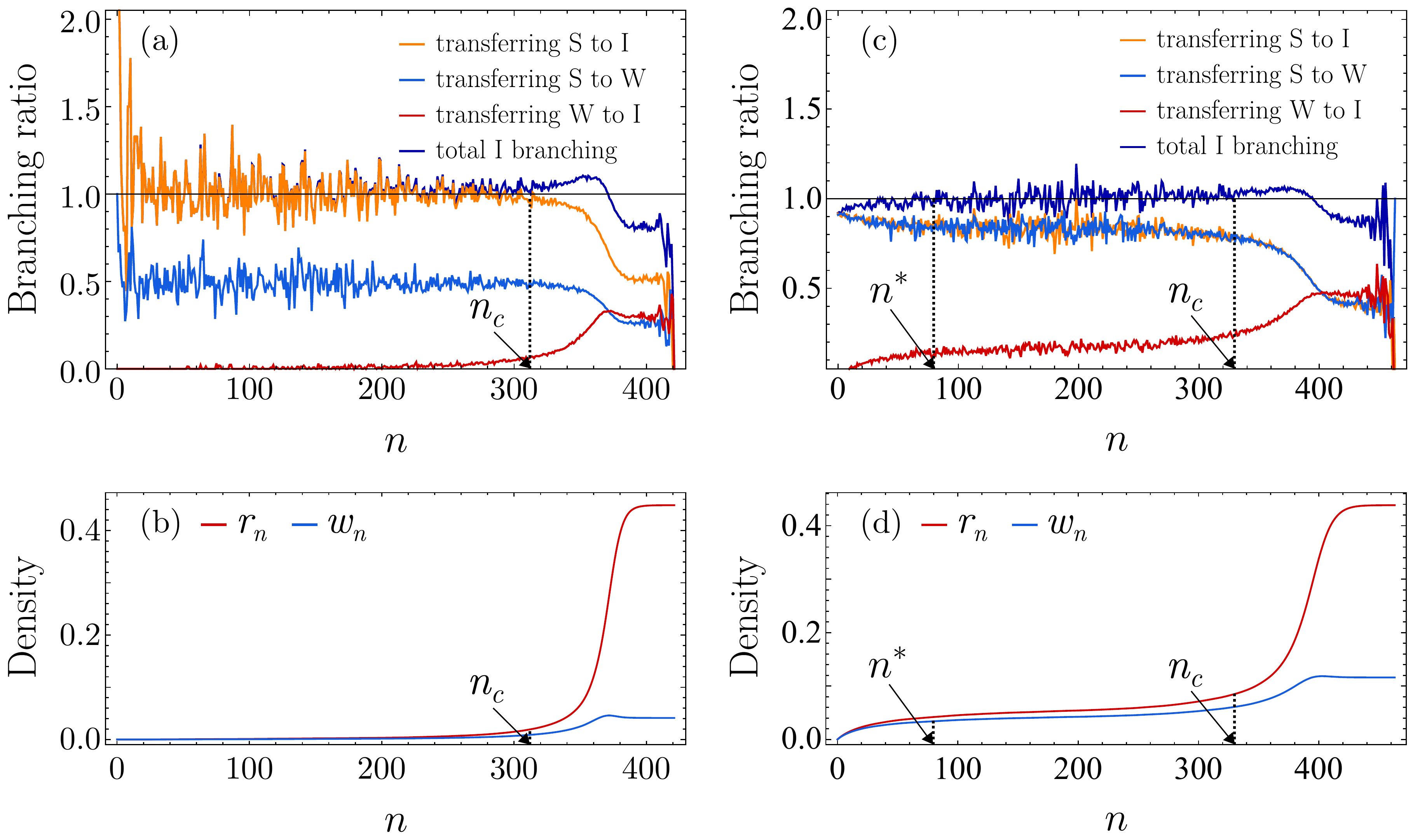}
	\caption{(a) For the single-seed case, plot of the branching ratios as a function of time step $n$ for each type of reaction at $\kappa_c =1/8$.  
		(b) Evolution of the densities of recovered nodes (red curve, top) and weakened nodes (blue curve, bottom) as a function of $n$ for the single-seed case. 
		Data are obtained from a single realization of infinite outbreak in the SWIR model starting from a single seed ($i_0=0$) with reaction probabilities $\mu=1/16$ and $\eta=0.9$ for both (a) and (b). (c) Similar to (a) but for the multiple-seed case at $\kappa_c \approx 0.1149487$. 
		(d) Similar to (b) but for the multiple-seed case. For both (c) and (d), data are obtained using the parameters $i_0=0.002$, $\mu=\kappa$, and $\eta=0.5$. The ER networks on which the simulations were performed have a size $N=5.12 \times 10^6$ and mean degree $z=8$. Legends ``transferring $X$ to $Y$'' in (a) and (c) indicate the mean number of neighbors of an infected node that change their state from $X$ to $Y$ at step $n$. In (d), the characteristic time steps $n^*$ and $n_c$, from and at which the CB process starts and ends, respectively, are marked.}
	\label{fig:branching}
\end{figure*}

We pick up the contribution of the reaction $S+I\to 2I$ from Eq.~(\ref{eq:f(r_n)}) but neglect the contribution of the reaction $W+I\to 2I$. Then, the probability that a node becomes directly infected by $\ell$ infectious neighbors, which is denoted by $ \pr^{(S \to I)}(\ell) $, is given as 
\begin{equation}\label{prone}
\pr^{(S\to I)}(\ell) =\sum_{m=0}^{\ell-1}(1-\kappa-\mu)^{m}\kappa = \dfrac{\kappa}{\kappa+\mu}[1-(1-\kappa-\mu)^{\ell}].
\end{equation}
Applying the formula for the Poisson degree distribution to Eq.~(\ref{eq:sce}), we obtain that 
\begin{eqnarray}\label{directInfection}
F^{(S\to I)}(u_{n}) =\dfrac{\kappa}{\kappa+\mu}\Big[1-e^{-(\kappa+\mu) z u_{n}} \Big].
\end{eqnarray}
Because the order parameter increases from $m=0$, we assume that $u_n$ is small in the early time regime. Thus,  
\begin{equation}
u_{n+1}^{(S\to I)}= z \kappa u_{n}-a u_n^2 + O(u_n^3),
\label{eq:map_si}
\end{equation}
where $a\equiv \kappa(\kappa+\mu)z^2/2$.
Actually, the coefficient $z \kappa $ of the first-order term is the mean branching ratio in the early time regime. When the critical branching (CB) process occurs, the mean branching ratio becomes unity, so the transition occurs at $\kappa_c=1/z$. On the other hand, the discrete mapping (\ref{eq:map_si}) at $\kappa_c$ may be rewritten in the form of a saddle-node bifurcation, ${\dot u}^{(S \to I)}=-au^2$, where $u$ is a function of the continuous time variable $n$ and the overdot denotes differentiation with respect to it. Because $a > 0$, $u^*=0$ is a stable fixed point for $u \ge 0$, and this point represents the fixed point of the SIR model, indicating a second-order transition. 

Next, we consider the two successive reactions $S+I\to W+I$ and $W+I\to 2I$, in which a susceptible node becomes infected in two steps and eventually recovers. Because a node can be infected either by the reaction $S+I\to 2I$ or by the reactions $S+I\to W+I$ and $W+I\to 2I$,
the probability $f^{(S\to W \to I)}(u_n)$ can be obtained using the relation 
\begin{equation}
F^{(S\to W \to I)}(u_n)=F(u_n)-F^{(S\to I)}(u_n)
\end{equation}
as 
\begin{eqnarray}\label{ftwo}
F^{(S\to W \to I)}(u_n)&=&\dfrac{\mu}{\kappa+\mu}\Big[1+\dfrac{\eta}{\kappa+\mu-\eta} e^{-(\kappa+\mu)z u_n}\Big] \nonumber \\ 
&-& \dfrac{\mu}{\kappa+\mu-\eta}e^{-\eta z u_n}.
\end{eqnarray}
Again, using $ u_n \ll 1 $, we obtain that 
\begin{eqnarray}\label{eq:map_swi}
u_{n+1}^{(S\to W\to I)}= b u_n^2+O(u_n^3),  
\end{eqnarray}
where $b\equiv {\mu \eta}z^2/2$.
Here we note that the first-order term $O(u_n)$ is absent. 
%This discrete mapping is rewritten in the form of a transcritical bifurcation, ${\dot u}=-u+bu^2$. There exist one stable fixed point $u^*=0$ and one unstable fixed point $u^*=1/b$. For $0< u < 1/b$, the dynamics flows into $u^*=0$, whereas for $u > 1/b$, it flows out from $u^*=1/b$. 
Combining Eqs.~(\ref{eq:map_si}) and (\ref{eq:map_swi}), we obtain that 
\begin{equation}\label{eq:map_total}
u_{n+1}=u_n+(b-a)u_n^2+O(u_n^3).
\end{equation} 
Thus, ${\dot u}=(b-a)u^2$. When $b-a < 0$, i.e., $\mu \eta < \kappa_c^2+\kappa_c \mu$, the fixed point $u^*=0$ is stable, and thus a continuous transition occurs. Otherwise, the fixed point $u^*=0$ is unstable, and a discontinuous transition occurs. The condition $\mu \eta > \kappa_c^2+\kappa_c \mu$ for a discontinuous transition is consistent with previously obtained results~\cite{grassberger_pre,choi_2016}.

When contagion starts from a single infectious node, its spread in the early time regime is governed by the linear term of Eq.~(\ref{eq:map_total}). It proceeds in the form of a CB tree~\cite{universal}, i.e. the mean branching ratio $(u_{n+1}-u_n)/(u_n-u_{n-1})$ is almost unity, and the main contribution is that of the reaction $S+I\to 2I$.
Thus in the thermodynamic limit, $u_n$ always stays zero so that nonlinear terms in Eq.~(\ref{eq:map_total}) do not appear. On the other hand, in finite systems, $u_n$ grows gradually and the nonlinear term $(b-a)u_n^2$ becomes significant after a characteristic time $n_c(N)$. It was argued in \cite{bennaim} that for the SIR model at the epidemic threshold, the maximum size of outbreaks is proportional to $N^{2/3}$ in the mean field limit. 
When $u_n$ grows up to $O(N^{2/3})$, the nonlinear terms in Eq.~(\ref{eq:map_si}) suppresses further growth of the cluster, leading to a subcritical branching process. This means that the CB process driven by Eq.~(\ref{eq:map_si}) persists up to $O(N^{1/3})$, because the fractal dimension of the CB tree is two. On the other hand, for the SWIR model, the coefficient of the nonlinear term (\ref{eq:map_total}) is positive, and the nonlinear term enhances further increase of removed nodes. The CB process turns into a supercritical process, leading to a pandemic outbreak. Accordingly, the golden time, the duration of the CB process, scales similarly as $\sim N^{1/3}$ to that of the SIR model, which is what we observed in a previous work~\cite{choi_2016,universal}.

\subsection{The multiple-seed case}

Next, when the number of infectious nodes is $O(N)$, i.e., $i_0>0$, $s_0=1-i_0$, and $w_0=0$ in the thermodynamic limit, it was shown~\cite{janssen_spinoal,hasegawa1,hasegawa3,choi_multi} that there exists a critical value $i_0^{(c)}$ such that when $i_0 < i_0^{(c)}$, a hybrid phase transition occurs at a transition point $\kappa_c$, whereas when $i_0=i_0^{(c)}$, a continuous transition occurs. Here we focus on the former case.

In the multiple-seed case, Eq.~(\ref{eq:sce}) becomes
\begin{equation}\label{eq:f(r_n)_multi}
u_{n+1}=i_0 + (1-i_0)F(u_n)
\end{equation}
when the network is an ER network with mean degree $z$. Fixed points of Eq.~(\ref{eq:f(r_n)_multi}) satisfy the equation,  
\begin{equation}\label{eq:G}
G(u)\equiv i_0+(1-i_0)F(u)-u=0
\end{equation}
and the smallest solution among them is the order parameter $m(\kappa)$. We note that 
$G(u)$ contains the parameters $\kappa$, $\mu$, $\nu$, $z$ and $i_0$. 
As already shown in the single-seed case, for appropriately given values of $\mu$ and $\eta$, $G(0)=i_0$, $G'(0)=(1-i_0)(z\kappa -1)$ and $G''(0)=(1-i_0)(b-a)>0$. Thus when $i_0$ is sufficiently small, i.e., $i_0<i_0^{(c)}$, $m(\kappa)$ satisfies $G'\big(m(\kappa)\big)<0$ for values of $\kappa$ near zero. Then $m(\kappa)$ increases continuously as $\kappa$ is increased untill $\kappa$ reaches a critical value $\kappa_c$ such that $G\big(m(\kappa_c)\big)=G'\big(m(\kappa_c)\big)=0$ and $G''\big(m(\kappa_c)\big)>0$ are satisfied. We note that $\kappa_c$ depends on $i_0$ and $z$. When $i_0=0$, $\kappa_c$ is reduced to $1/z$, the transition point of the single-seed case. $m(\kappa)$ exhibits a critical behavior as $\kappa$ approaches $\kappa_c$ and subsequently jumps from $m(\kappa_c)=m_d$ to another value $m_u$ as represented in Fig.~\ref{fig:op}(b). Thus the transition is hybrid.

We notice that at a transition point for the multiple-seed case, an infected node can be in contact with a node that was weakened by a different infectious root~\cite{choi_multi}. Accordingly, the reaction $W+I\to 2I$ can occur even in the early time regime, as shown in Fig.~\ref{fig:branching}(c) with red zig-zag (lowest) curve. Moreover, the CB process appears not from the beginning but slightly after that indicated by an arrow at $n^*$ in Fig~\ref{fig:branching}(c), at which the density of recovered nodes $r_{n^*}$ is close to $m_d$ indicated in Fig.~\ref{fig:op}(b).
%\ynew{at which the total mean branching ratio of an infected node fluctuates around one and} the density of recovered nodes $r_{n^*}$ is close to $m_d$ indicated in Fig.~\ref{fig:op}(b). 
From this step $n^*$, $r_n$ remains almost constant for a long time as shown in Fig.~\ref{fig:branching}(d). 

%This means that $r_n$ is trapped around $m_d$ by the so-called bottle-neck effect as we discuss later. For some configurations, the dynamics is trapped at $m_d$ forever, but for other configurations, it escapes from the trap and so the  plateau ends at $n_c$, and then $r_n$ increases abruptly and reaches a stable point $m_u$. We focus on the latter case in the following.

However, in finite systems, due to the fluctuations arising in the stochastic process of epidemic spread, the densities of each species of nodes at $n^*$ can be different for each realizaton. Those fluctuations affect $n_c$, which can be also different for different realizations, where $n_c$ is the golden time, from which $r_n$ increases drastically.

%The recursive equation for $u_n$ is written as  
%\begin{eqnarray}\label{eq:multi_r_n}
%u_{n+1}=i_0+(1-i_0)f(u_{n},\kappa), 
%\end{eqnarray}
%where $f(u_n,\kappa)$ is \xnew{the same} to $f(u_n)$ in Eq.~(\ref{eq:f(r_n)}), but for further discussion, we explicitly present $\kappa$ in the form of $f(u_n,\kappa)$. 
%The epidemic dynamics begins at $n=0$ with the densities of susceptible, weakened, infected and recovered nodes $s_0=1-i_0$, $w_0=0$, $i_0=i_0$ and $r_0=0$, respectively. 
We denote the densities of each species of nodes at a certain time step $\ell$ as $s_{\ell}, w_{\ell},i_{\ell},$ and $r_{\ell}$, respectively. Then on ER networks, for $n > \ell$, $h_{n,\ell}\equiv u_n-r_{\ell}$ satisfies
\begin{equation}\label{eq:gre}
	h_{n+1,\ell}=i_{\ell}+\sum_{q=0}^{\infty}\dfrac{z^qe^{-z}}{q!}
	\Big(s_{\ell}f_{q}(h_{n,\ell})+w_{\ell}g_{q}(h_{n,\ell})\Big),
\end{equation}
where
\begin{align}\label{key}
s_\ell=s_0 e^{-(\kappa+\mu)zu_{\ell-1}}
\end{align}
and 
\begin{align}\label{key2}
w_\ell=1-u_\ell-s_\ell.
\end{align}
Moreover, using Eq.~(\ref{eq:sce}), the relation $i_n=u_{n+1}-u_{n}$, and $u_n=i_n +r_n$, we determine $i_\ell$ and $r_\ell$.

We focus on the density fluctuations of each species at $n^*$.  We split the densities of each species of nodes into two parts: $x_{n^*}+\delta x_{n^*}$ ($x=s,w,i$ or $r$), where the first term represents densities of $x$-species nodes in thermodynamic limit at $n^*$, and the second one is the deviation. Then for $n > n^*$, Eq.~(\ref{eq:gre}) becomes
\begin{eqnarray}\label{eq:gre_N}
h_{n+1,n^*}&=&{i}_{n^*}+\delta i_{n^*}+({s}_{n^*}+\delta s_{n^*})f(h_{n,n^*}-\delta r_{n^{*}}) \nonumber \\
&&+({w}_{n^*}+\delta w_{n ^*})g(h_{n,n^*}-\delta r_{n^{*}})+\delta r_{n^*}.
\end{eqnarray}
We did not take into account the density fluctuations induced after $n^*$, because they are negligible compared to those at $n^*$.
At $\kappa=\kappa_c$, Eq.~(\ref{eq:gre_N}) has a nontrivial fixed point $h_{d,n^*}=m_d-r_{n^*}$ in the thermodynamic limit. Then Eq.~(\ref{eq:gre_N}) is rewritten with $\epsilon_n=u_n-m_d$ as
\begin{equation}\label{eq:e_n2}
\epsilon_{n+1}=d_0+(1+\delta d_1)\epsilon_n+(d_2+\delta d_2) \epsilon_n^2+O(\epsilon_n^3),
\end{equation}
where
\begin{widetext}
\begin{eqnarray}
d_0 &\approx& \delta i_{n^*}+\delta s_{n^*}f(h_{d,n^*})+\delta w_{n^*}g(h_{d,n^*}),\label{key} \\
\delta d_1 &\approx& \delta s_{n^*}f'(h_{d,n^*})+\delta w_{n^*}g'(h_{d,n^*})-\big(s_{n^*}f''(h_{d,n^*})
+ w_{n^*}g''(h_{d,n^*})\big) \delta r_{n^{*}}, \label{key} \\
d_2 &=& \dfrac{1}{2}\big( s_{n^*}f''(h_{d,n^*})+w_{n^*}g''(h_{d,n^*}) \big),\label{key} \\
\delta d_2 &=& \dfrac{1}{2}\big( \delta s_{n^*}f''(h_{d,n^*})+\delta w_{n^*}g''(h_{d,n^*}) \big) -\dfrac{1}{2}\big(s_{n^*}f'''(h_{d,n^*})+w_{n^*}g'''(h_{d,n^*})\big) \delta r_{n^{*}}.\label{key}
\end{eqnarray}
\end{widetext}
Neglecting higher order terms of $\epsilon$, Eq.~(\ref{eq:e_n2}) is rewritten in an alternative form, 
\begin{equation}\label{eq:bifurcation}
\dot{\epsilon}=d_0+(d_2+\delta d_2)\bigg(\epsilon+\dfrac{\delta d_1}{2(d_2+\delta d_2)}\bigg)^2
-\dfrac{(\delta d_1)^2}{4(d_2+\delta d_2)}.
\end{equation}
Because $\delta i_{n^*}\sim \delta s_{n^*}\sim\delta w_{n^*}\sim\delta r_{n^*} \ll 1$ for large $N$, the last term can be neglected compared to $d_0$ and Eq.(\ref{eq:bifurcation}) is rewritten simply as 
\begin{equation}\label{eq:bifurcation2}
\dot{\epsilon'}=d_0+{d_2}'\epsilon'^2
\end{equation}
where ${d_2}'=d_{2}+\delta d_2$ and $\epsilon'=\epsilon+\delta d_1/2{d_2}'$. We note that $d_0$ is a real number, while $d_2^{\prime}$ is a positive number.   
%where $d_1\sim O(1/\sqrt{N})$.

\begin{figure}[t]
	\centering
	\includegraphics[width=1.0\linewidth]{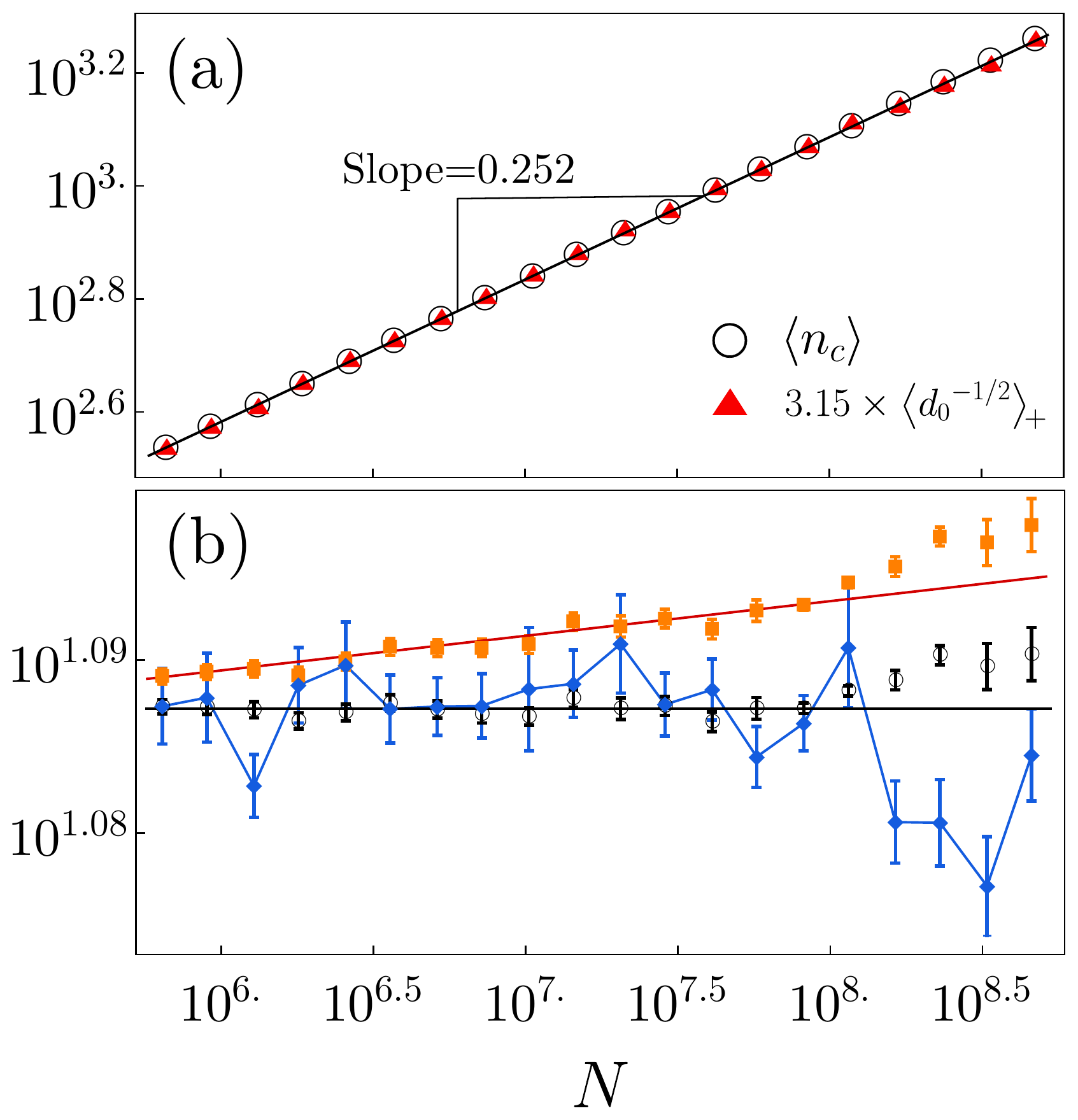}
	\caption{For the multiple-seed case, (a) plot of the average cascade time step $\langle n_c \rangle$ ($\bigcirc$) and $3.15\big\langle {{d_0}^{-1/2}}\rangle_+$ (\textcolor{red}{$\blacktriangle$}) versus system size $N$ at $\kappa_c \approx0.11494875096512$. The notation $_+$ in $\big\langle {d_0}^{-1/2}\rangle_+$ indicates that only positive values of $d_0$ are considered in taking the average. $d_0$ is measured at $n^*=80$. The guideline has a slope of $0.252$.  (b) Plot of $\langle n_c \rangle / N^{0.25}$ (\textcolor{orange}{$\blacksquare$}), $1.023\langle n_c \rangle / N^{0.252}$ ($\bigcirc$), and $3.3\dsqrt /N^{0.254}$ (\textcolor{blue}{$\blacklozenge$}) versus $N$. Data were obtained from ER networks of different sizes but with the same mean degree, $z=8$. $i_0=0.002$, $\mu=\kappa$, and $\eta=0.5$ were used. Average is taken over more than $10^5$ realizations for each data point for $N < 10^8$.}
	\label{fig:SWIRGoldenTime}
\end{figure}

\begin{figure*}
	\centering
	\includegraphics[width=1.0\linewidth]{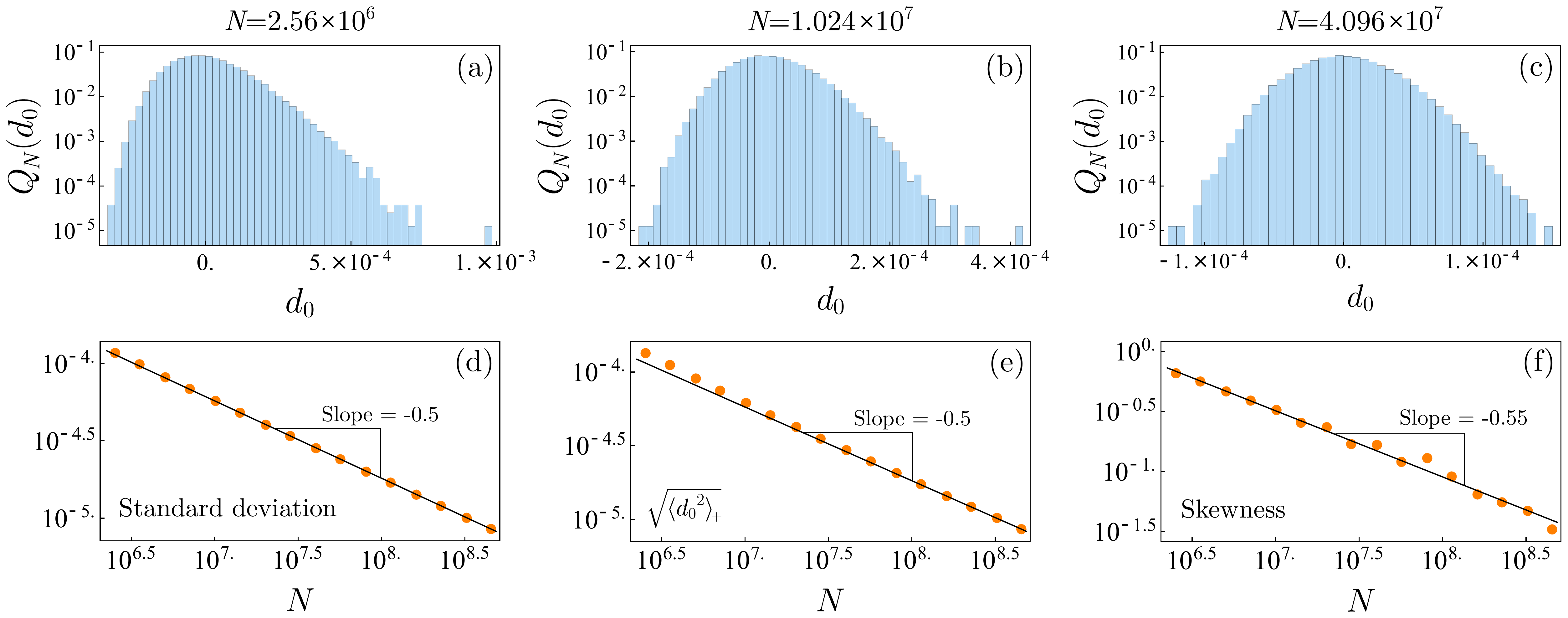}
	\caption{Plot of the probability distribution $Q_N(d_0)$ versus $d_0$ for different system sizes, $N=2.56\times 10^6$(a), $1.024\times 10^7$ (b), and $4.096\times 10^7$ (c). The distribution is obtained from more than $10^5$ realizations for each $N$. Plot of the standard deviation (d), $\sqrt{\langle {d_0}^2\rangle_{+}}$ (e), and the skewness (f) of $Q_N(d_0)$ as a function of the system size $N$. Average is taken over more than $10^5$ realizations for each data point.}
	\label{fig:qd1}
\end{figure*}
%\begin{figure}
%	\centering
%	\includegraphics[width=1.0\linewidth]{positiveRatio.pdf}
%	\caption{Plot of fracion of positive $d_1$ as a function of the system size $N$.}
%	\label{fig:s_moment_plus}
%\end{figure}

The nonlinear mapping Eq.(\ref{eq:bifurcation2}) includes several features: When $d_0 < 0$, $\epsilon^{\prime}$ reaches 
%to 
a fixed point 
$\epsilon_*^{\prime}=-\sqrt{|d_0|/d_2^{\prime}}$; when $d_0=0$, $\epsilon^{\prime}$ remains at zero; when $d_0 > 0$, there arises the so-called bottleneck effect at $\epsilon^{\prime}=0$~\cite{book,kcore3,buldyrev}. The time step to pass through the bottleneck is calculated as 
\begin{equation}\label{eq:bottleneck}
{\cal{T}}=\int_{-\infty}^{\infty} \frac{d\epsilon'}{d_0+{d_2}'\epsilon'^2}\sim \frac{\pi}{\sqrt{d_0}},
\end{equation}
which is approximately the time interval of the plateau region, i.e, $n_c-n^*$. Because $n^*$ is much smaller than $n_c$, $n_c\approx \cal{T}$, which is the golden time for a single realization of the process. $d_0$ can have different values for different realizations, yielding a different $n_c$. Thus, we need to take average of $n_c$ over different realizations to obtain $\langle n_c \rangle$.

We performed extensive numerical simulations at the transition point $\kappa_c \approx0.11494875096512$ of the SWIR model starting from multiple seeds $i_0=0.002$, and obtained that 
\begin{equation}
\langle n_c \rangle \sim N^{0.252\pm 0.001}
\label{eq:n_c}
\end{equation}
and
\begin{equation}
\big\langle d_0^{-1/2} \big\rangle_+ \sim N^{0.254 \pm 0.002}
\label{eq:scaling_d_1_2}
\end{equation} 
as shown in Fig.~\ref{fig:SWIRGoldenTime}. $\langle \cdots \rangle_+$ represents the ensemble average over only positive values of $d_0$. Otherwise, $\epsilon^{\prime}$ does not diverge by repeating iterations. We remark that the exponent value is larger than $1/4$. The numerical exponent values in Eqs.~\eqref{eq:n_c} and \eqref{eq:scaling_d_1_2} are obtained with the data only within the range $N < 10^8$. The data beyond that range get out of the trend abruptly, which may be caused by too long passing time through too narrow bottle necks as the system size becomes large. The noise term $d_0$ was obtained at $n^*\approx 80$ in Fig.~\ref{fig:branching}, at which a critical branching process starts. 

Now we consider the distribution of $d_0$ obtained from different realizations but at the same $n^*$ for the system size $N$, denoted as $Q_N(d_0)$. We define the standard deviation $\sigma_N$ of $Q_N(d_0)$ as  
$$
\sigma_N^2=\langle d_0^2 \rangle -\langle d_0 \rangle^2,
$$ 
where $\langle \cdots \rangle$ represents the average over all range of $d_0$ and $\langle d_0 \rangle > 0$. 
$\sigma_N$ behaves as $\sim N^{-1/2}$ as shown in Fig.~\ref{fig:qd1}(d). If we assume that any moment of the distribution $Q_N(d_0)$ is determined by the single scale, so that 
$$\dsqrt  \sim \big\langle d_0^2 \big\rangle_+^{-0.25} \sim \sigma_N^{-1/2},$$ 
then it would behave as $N^{1/4}$. However, this result is not consistent with the numerical result \eqref{eq:scaling_d_1_2}. 

We check the $N$-dependent behavior of $\sqrt{\langle d_0^2 \rangle_+}$. Fig.~\ref{fig:qd1}(e) shows that $\sqrt{\langle d_0^2 \rangle_+}$ behaves as $N^{-1/2}$ asymptotically but the data points deviate in small $N$ region. This discrepancy mainly originates from the asymmetry of $Q_N(d_0)$, which is caused by the multiplicative noise induced by the stochastic process. $Q_N(d_0)$ has a longer tail in its positive side than in the oppposite side as shown in Fig.~\ref{fig:qd1}(a)$-$(c). As the system size becomes larger, it becomes not only narrower but also more symmetric. To quantify this asymmetric feature of the distribution, we measure the skewness of $Q_N(d_0)$ defined as 
\begin{equation}
S_3 \equiv \Bigg\langle\Bigg(\frac{d_0-\langle d_0 \rangle}{\sigma_N}\Bigg)^3\Bigg\rangle \sim N^{-0.55},
\end{equation}
in Fig.~\ref{fig:qd1}(f). The above result suggests that the distribution remains asymmetric in any finite systems but becomes symmetry only in the limit $N\to \infty$. 
$Q_N(d_0)$ becomes a Gaussian distribution in that limit. 
The asymetry of $Q_N(d_0)$ decreases because the ratio of the noise to the mean number of infected nodes becomes smaller for larger systems.

Due to those features, $\dsqrt$ behaves differently from $\sigma_N ^{-1/2}$ within our numerical range; however, it is not certain yet how it would be in the thermodynamic limit because our simulation data (Fig.~\ref{fig:SWIRGoldenTime}) of $\dsqrt$ contain heavy fluctuations, particularly in the large-system-size region. For much larger system sizes, $Q_N(d_0)$ are so close to the Gaussian distribution that one may think that $\dsqrt$ behaves as $\sigma_N ^{-1/2}$, i.e., $\sim N^{1/4}$ in the thermodynamic limit $N\to \infty$. 
However, it is a challenging task to verify that numerically. 

\begin{figure}[t]
\centering
\includegraphics[width=1.0\linewidth]{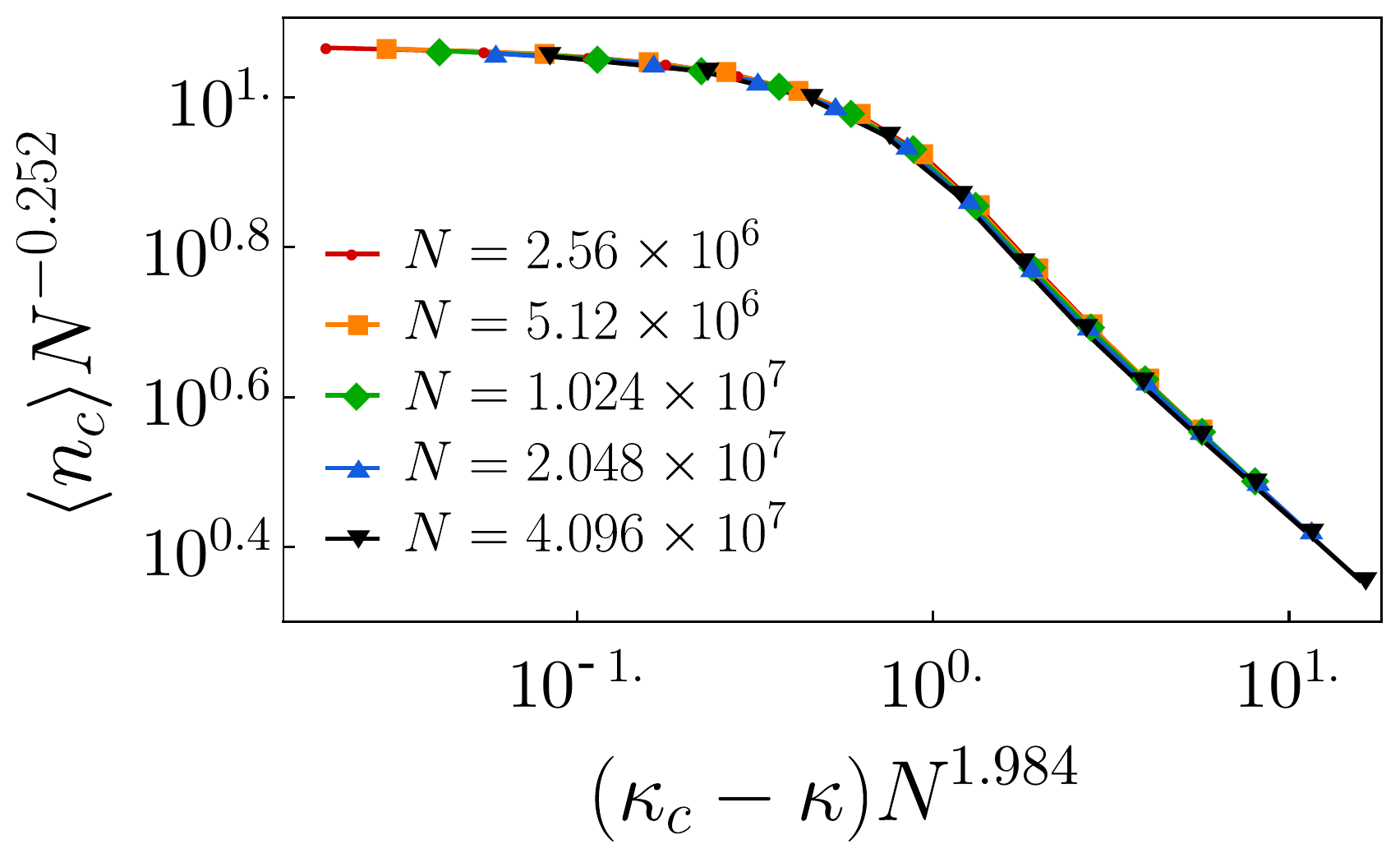}
\caption{For the multiple-seed case, scaling plot of the average cascade time step $\langle n_c \rangle N^{-0.252}$ versus $(\kappa-\kappa_c)N^{1.984}$ for different system sizes $N$. Data for different system sizes collapse well onto a single curve, indicating that $\langle n_c \rangle \sim N^{1/4}$ for $\kappa > \kappa_c$. Numerical simulations were performed on ER networks with mean degree $z=8$ and initial density of seeds $i_0=0.002$. Average is taken over more than $5\times 10^4$ realizations for each data point.}
\label{fig:sqrtLaw}
\end{figure}

When $\kappa > \kappa_c$, $d_0$ is naturally obtained as 
$d_0=(1-i_0)({{\partial f(u_n, \kappa)}/{\partial \kappa}})\big|_{m_d,\kappa_c}(\kappa-\kappa_c)$. 
Then, we do not need to take average over ensembles for sufficiently large $\kappa-\kappa_c$ because sample to sample fluctuations of $d_0$ become negligible compared to it. Then, 
\begin{equation}\label{eq:bottleneck_theory}
\langle n_c \rangle=\int_{-\infty}^{\infty} \frac{d\epsilon}{d_0+d_2\epsilon^2}\sim \dfrac{\pi}{\sqrt{\kappa-\kappa_c}}.
\end{equation} 
Numerical result in Fig.~\ref{fig:sqrtLaw} supports this prediction. 

\begin{figure}[t]
\centering
\includegraphics[width=1.0\linewidth]{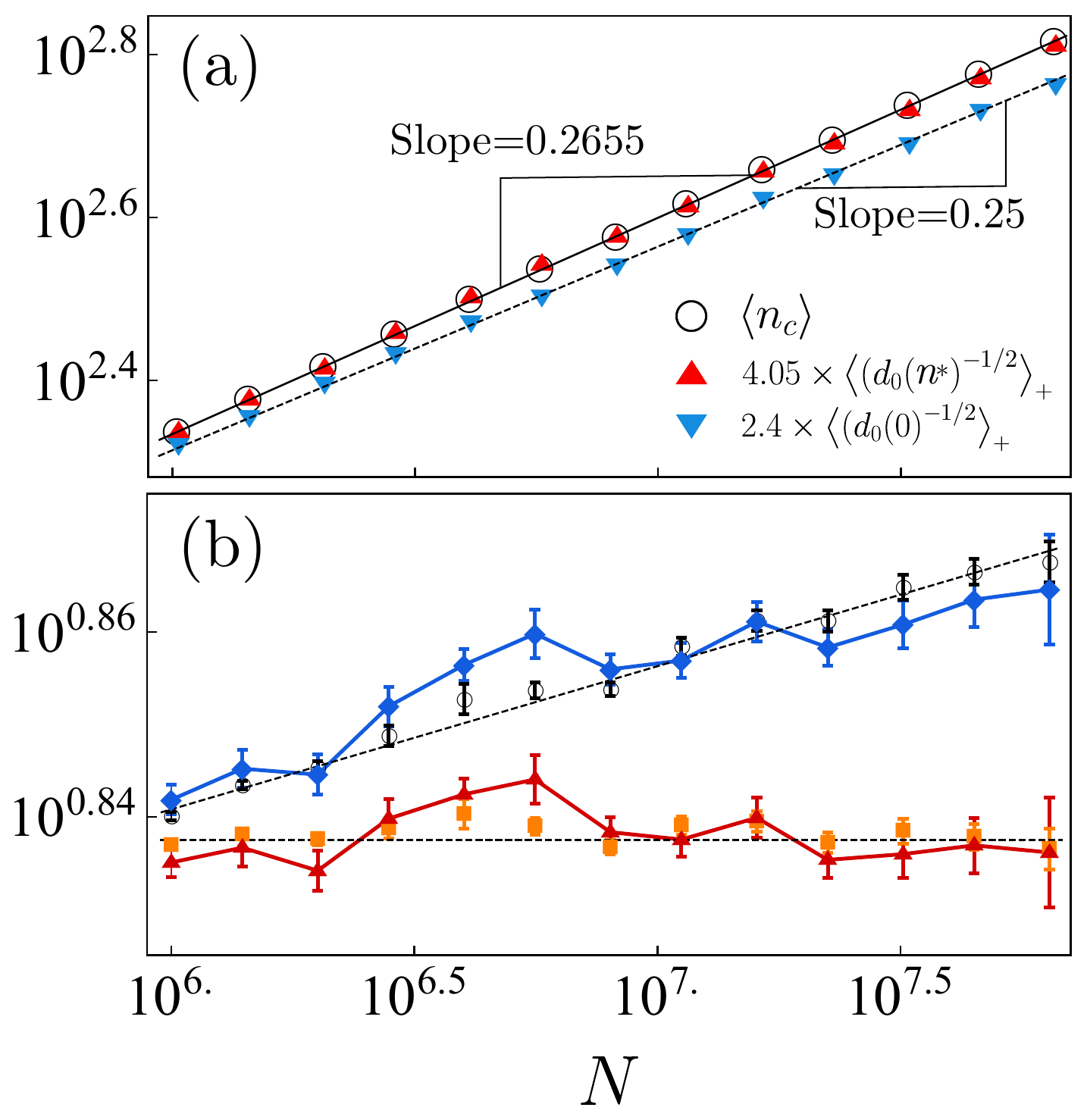}
\caption{For $k$-core percolation with $k=3$, (a) plot of the golden time $\langle n_c \rangle$ ($\bigcirc$) and  $4.05\big\langle {d_0(n^*)}^{-1/2}\big\rangle_+$ (\textcolor{red}{$\blacktriangle$}), and $2.4\big\langle {d_1(0)}^{-1/2}\big\rangle_+$ (\textcolor{blue}{$\blacktriangledown$}) versus $N$ for $k$-core percolation with $k=3$ starting from multiples nodes of $O(N)$. $d_0$ is measured at $n^*=60$ and $d_0(0)$ denotes the value of $d_0$ measured at $n=0$. A solid (dashed) guideline has slope of 0.2655 (0.25). (b) Plot of $1.23\langle n_c\rangle/N^{0.2655}$ (\textcolor{orange}{$\blacksquare$}), $\langle n_c\rangle/N^{0.25}$ ({$\bigcirc$}), $4\big\langle {d_0(n^*)}^{-1/2}\big\rangle_+/N^{0.25}$ (\textcolor{blue}{$\blacklozenge$}), and $4.65\big\langle {d_0(n^*)}^{-1/2}\big\rangle_+/N^{0.262}$ (\textcolor{red}{$\blacktriangle$}) versus $N$. Data were obtained from ER networks of different sizes $N$ but with the same mean degree, $z=3.723243$, and $i_0= 0.0567377$. Average is taken over more than $10^5$ realizations for each data point.}
\label{fig:kcore}
\end{figure}

\begin{figure}[t]
\centering
\includegraphics[width=1.0\linewidth]{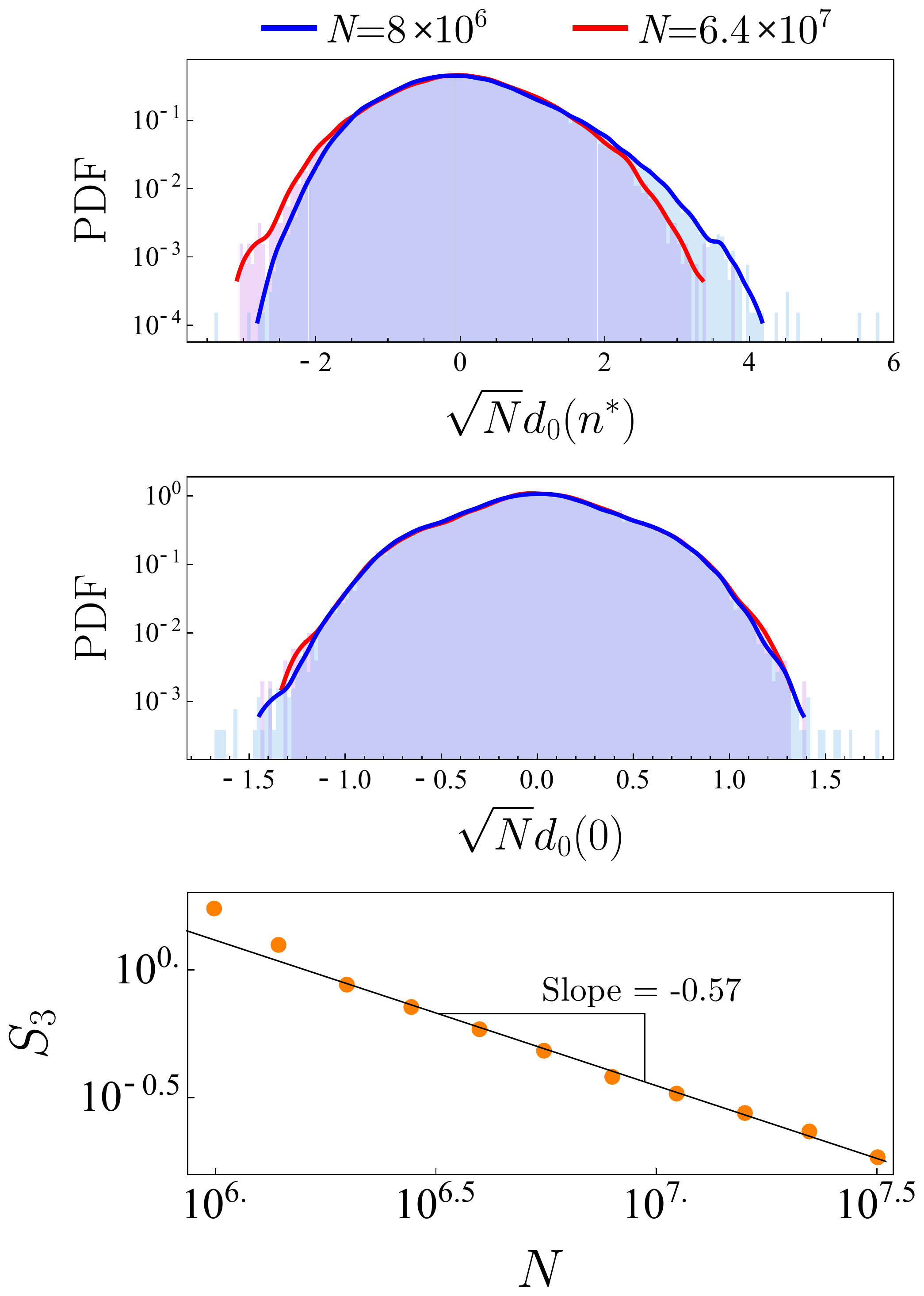}
\caption{Plot of the probability density function of (a) $\sqrt{N}d_0(n^*)$ and (b) $\sqrt{N}d_0(0)$ for two different system sizes $N=8\times 10^6$ and $6.4\times 10^7$. (c) Plot of skewness $S_3$ of $Q_N(d_0(n^*))$ as a function of the system size $N$. Average is taken over more than $10^6$ realizations for each data point.}
\label{fig:pdf}
\end{figure}

\begin{figure}[t]
\centering
\includegraphics[width=1.0\linewidth]{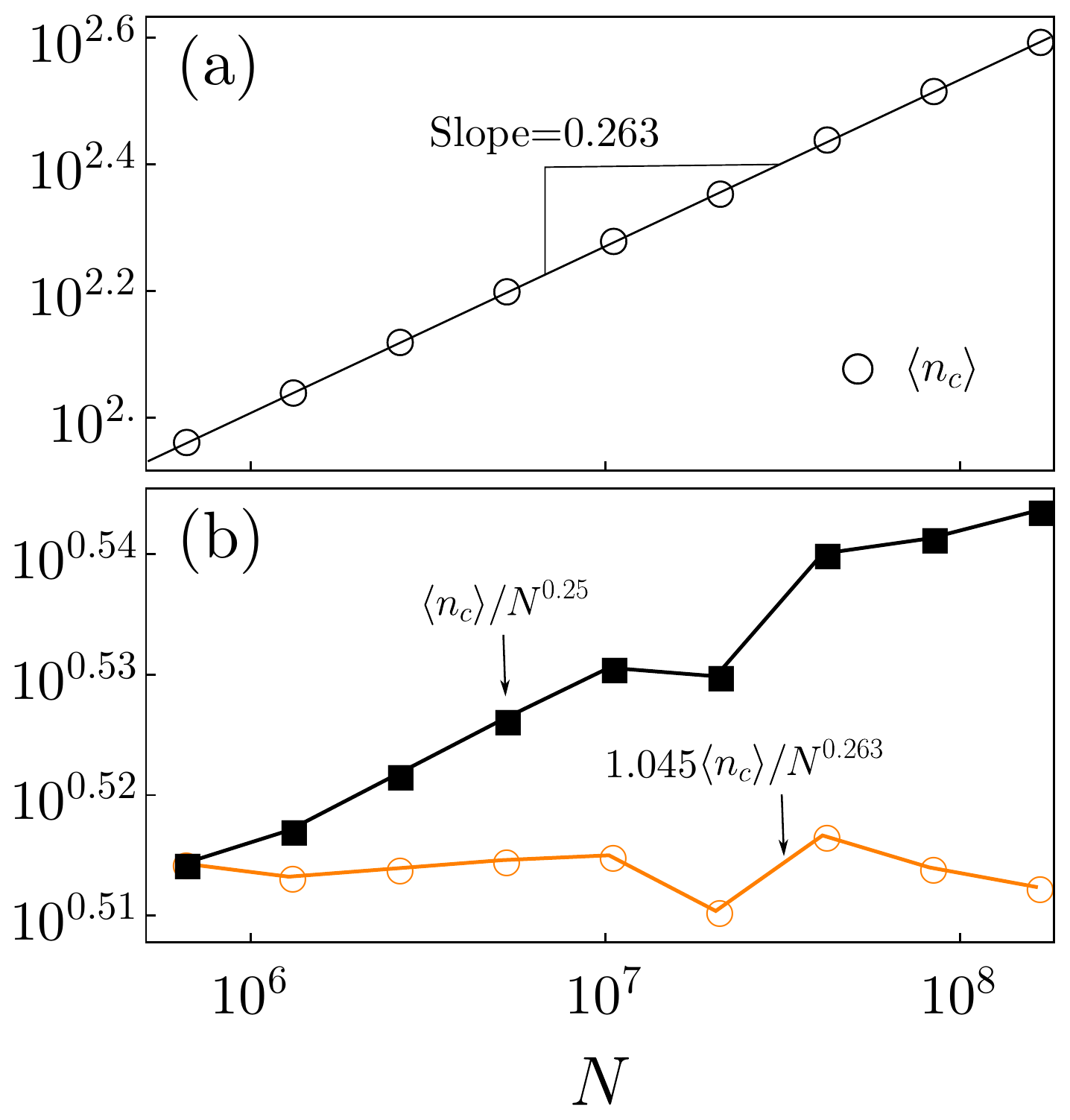}
\caption{(a) Plot of the average cascade time step $\langle n_c \rangle$ for the threshold model starting from initial multiple active nodes. Guideline has a slope of $0.263$. (b) Plot of $\langle n_c\rangle/N^{0.25}$  ($\blacksquare$) and $\langle n_c\rangle/N^{0.263}$ (\textcolor{orange}{$\bigcirc$}) versus $N$. Data were obtained from ER networks of different sizes $N$ but with the same $(z,\rho_0,\phi)=(9.191,0.01,0.18)$. Average is taken over more than $4\times 10^4$ realizations for each data point.}
\label{fig:threshold}
\end{figure}

\section{$K$-core percolation and the threshold model}

In our previous work~\cite{universal}, we showed that there exists universal mechanism of avalanche dynamics in the SWIR model, $k$-core percolation, the threshold model and the CFoIN, when an avalanche starts from a single seed. Due to the universal mechanism, the golden time scales as $N^{1/3}$ in a universal way for those models. During that study, we found that numerical simulations for the CFoIN with large system sizes require long computational times and memory space, so that numerical results with limited ensemble average were not neat. Based on such experience, here we limit our interest on to the golden time problem with multiple seeds to $k$-core percolation~\cite{kcore1} and the threshold model~\cite{watts} besides the SWIR to check the universal behavior. We find that for both models, the exponent values of the golden time are also measured to be slightly larger than 1/4. We note that the SWIR model and the two models above can be regarded as special cases of generalized epidemic process~\cite{grassberger_pre,dodds} with heterogeneous transmission probabilities. Thus, similar behaviors of golden time are expected. Let us begin with $k$-core percolation. 

\subsection{$k$-core percolation} 
Here we first consider the avalanche dynamics of $k$-core percolation. First we construct a $k$-core subgraph from an ER random graph with mean degree $z$. When $z$ is larger than a threshold $z_c$, a $k$-core subgraph of size $O(N)$ can exist. After this step, $\rho_0N$ nodes are removed simultaneously. There may exist some nodes that have degree less than $k$. In this case, those nodes are removed repeatedly until no more such nodes remain.  
The avalanche size can be either finite or infinite depending on $z$ and $\ro$. If it is finite, the $k$-core would still exist; If it is infinite, the $k$-core would collapse to zero. For sufficiently large $z$, there exists a critical density $\rho_c$ such that an infinite avalanche can occur when $\ro > \rho_c $ in the thermodynamic limit. In Fig.~\ref{fig:kcore}, we measure the mean cascade time step (golden time) $\langle n_c \rangle$ of infinite avalanches at the transition point for different system sizes $N$. We also measure $\big\langle {d_0(n^*)}^{-1/2}\big\rangle_+$ and $\big\langle {d_0(0)}^{-1/2}\big\rangle_+$, where $d_0(n^*)$ is the noise measured at $n^*=60$, at which a critical branching process occurs. The definition of $d_0$ is presented in Appendix A. $d_0(0)$ is measured at $n=0$, which represents structural fluctuation at $n=0$. It was found that $\langle n_c \rangle$ scales as  $N^{0.2655\pm0.003}$ and $\big\langle {d_0(n^*)}^{-1/2}\big\rangle_+$ scales as  $N^{0.262\pm0.007}$. On the other hand, $\big\langle {d_0(0)}^{-1/2}\big\rangle_+$ is proportional to $N^{0.25}$. 

The distribution functions of $\sqrt{N}d_0(0)$ and $\sqrt{N}d_0(n^*)$ are shown in Fig.~\ref{fig:pdf} for two different system sizes. Similar to the case of the SWIR model, $d_0(n^*)$ is distributed asymmetrically and the distribution becomes more symmetric for larger system sizes (Fig.~\ref{fig:pdf}(a)). Such factors make the exponent larger than 1/4. On the other hand, since the multiplicative fluctuations of cascade dynamics are absent at $n=0$, the distribution of $d_0(0)$ do not change in shape for different system sizes (Fig.~\ref{fig:pdf}(b)). Thus it satisfies $Q_N(d_0(0))=\sqrt{N}Q(\sqrt{N}d_0(0))$, which makes $\big\langle {d_0(0)}^{-1/2}\big\rangle_{+}$ scale as $\sim N^{1/4}$. However, it is also uncertain that the value $\zeta$ remains unchanged for larger systems.

\subsection{The threshold model}
Next we consider the threshold model, which was introduced to study the spread of cultural fads on social networks. Each node $i$ is assigned its threshold value $\phi_i$ and has one of two states, active or inactive. An inactive node $i$ surrounded by $m_i$ active neighbors and $k_i-m_i$ inactive neighbors changes its state to active when the fraction of active neighbors $m_i/k_i > \phi_i$. For a given set of $\{\phi_i\}$, the order parameter, the density of active nodes in an absorbing state, jumps and exhibits a hybrid phase transition at a critical value of the mean degree $z_c$.
Here, we initially introduce $i_0 N$ active nodes in a system. At each generation, every inactive node $i$ whose number of active neighbors $m_i > k_i \phi_i$ is identified and changes its state to active. For convenience, we choose a single threshold value $\phi$ for all nodes on ER networks. Then the critical mean degree $z_c$ is determined as a function of $\phi$ and $i_0$. We performed simulations with $\phi=0.18$ and  $i_0=0.01$. Then the critical point is determined as $z_c=9.191\dots$ in the thermodynamic limit. The mean cascade time step of infinite outbreaks, $\langle n_c \rangle$, is obtained numerically as $\sim N^{0.263}$ in Fig.~\ref{fig:threshold}. Thus the measured exponent is also larger than $1/4$.

\section{puzzle in CFoIN}

A similar size dependency of the golden time was addressed in the CFoIN. Zhou et al.~\cite{zhou} revealed that the choices of different types of transition points lead to different scaling behaviors of golden time in the CFoIN. They showed that when the golden time is measured at the transition point $p_c$ of each realization, the mean golden time scales as $\langle n_c \rangle \sim N^{1/3}$.
On the other hand, when a single mean-field transition point $p_c^{\textrm{MF}}$ is taken for all realizations, the golden time scales as $\langle n_c^\textrm{MF} \rangle  \sim N^{1/4}$. The authors presented the hand-waving  argument that in finite systems of size $N$, individual $p_c$ follows a standard Gaussian distribution having the mean value $p_c^{\textrm{MF}}$ and the standard deviation proportional to $N^{-1/2}$~\cite{buldyrev,zhou}. Using a formula similar to Eq.~\eqref{eq:bifurcation2} with $d_0$ following a Gaussian distribution having the standard deviation $\sim N^{-1/2}$, they obtained $\langle n_c^\textrm{MF} \rangle  \sim N^{1/4}$. On the other hand, the author of Ref.~\cite{grassberger_mcc} investigated the scaling relation of the golden time numerically using a different algorithm, and obtained the exponent $\zeta \approx 0.28$ different from $1/4$. Thus, the two results are not consistent with each other and this discrepancy has remained as a puzzle in the cascade-induced discontinuous percolation. 

We recall that for the SWIR model, $i_0N$ seeds are selected at random. Thus, the dynamics started from those nodes can be different for each sample. Because these choices are random, the distribution of $d_0$ at $n=0$ will follow a Gaussian distribution in a similar way to the $k$-core percolation case. However, because the dynamics proceeds from $n=0$ stochastically, noises are accumulated during the avalanche dynamic process. In this case, for a given network at $n=0$, noises of $d_0$ obtained at $n^*$ do not form a regular Gaussian distribution but do an asymmetric distribution, and the observed scaling of $\langle n_c \rangle$ is not $N^{1/4}$. We think that the result obtained in Ref.~\cite{grassberger_mcc} shares the common origin with the one we have in the SWIR model. Therefore, we think that the puzzle arising between the results of Refs.~\cite{zhou} and \cite{grassberger_mcc} originates from the times at which the distribution of the fluctuation is measured.

\section{Summary}

The SWIR model is a simple two-step contagion model, enabling us to understand the machanism underlying a pandemic outbreak. Using this model, we obtained the scaling behavior of the golden time with respect to the system size. Using the local tree approximation, we set up a nonlinear dynamic equation in the form of saddle-node bifurcation that represents the cascade dynamics of two-step contagion. When the epidemic dynamics starts from a single infected node, we showed that the linear and the nonlinear terms of the nonlinear mapping play distinct their roles. In the early time regime, the linear term governs a critical branching (CB) process. The CB tree can be regarded as a critical cluster in percolation. However, in the late time regime, the nonlinear term causes an explosive spread of epidemic disease. The golden time is determined by the finite-size effect on the linear term, which scales as $\sim N^{\zeta}$ with $\zeta=1/3$. This scaling behavior is universal for cascade-induced dynamic models such as the threshold model, $k$-core percolation and the CFoIN. 

When the dynamics starts from multiple seeds of $O(N)$, we measured a change in the value of $\zeta$ to 0.252 in the SWIR model. In this case, a long CB process does not appear from $n=0$, but it does at some characteristic time $n^*$. During the time until $n^*$, clusters of ever infected nodes merge and form a cluster of size $O(N)$. The size fluctuates for different realizations. 
The fluctuations are induced by the stochastic process of disease transmission. We found that these fluctuations change the value of $\zeta$ from 1/3 to about 0.252 for  the multiple-seed case. Due to the multiplicative noise of disease spread, the size distribution of those clusters over different realizations becomes asymmetric with a long tail in its positive region. It seems that due to such non-Gaussianity the golden time scales as $\sim N^{\zeta}$ with $\zeta$ slightly larger than $1/4$. However, this asymmetry decreases gradually in a power-law manner of the skewness function as the system size is increased. This leaves the possibility open that asymptotically the value of $\zeta$ approaches $1/4$. This problem could not be ultimately solved by our study. A very precize analysis of corrections to scaling would be needed for it, what was not possible in spite of our massive numerical efforts.
 
We also obtained the similar behavior, $\zeta > 1/4$ for the two other cascade dynamics models, $k$-core percolation and the threshold model. On the basis of the numerical results of the SWIR model, the threshold model and $k$-core percolation, the exponent $\zeta$ seems to be non-universal for the multiple-seed case. However, as the $\zeta$ exponents for those models deviate only slightly from 1/4 and the simulation  sizes are limited, the asymptotic universal behavior cannot be entirely excluded. \\

\begin{acknowledgments}
This work was supported by the National Research Foundation of Korea by grant no. NRF-2014R1A3A2069005 and by H2020 FETPROACT-GSS CIMPLEX Grant No. 641191 (JK).
\end{acknowledgments}

\appendix

%\hskip 1cm

\section{Derivation of $d_0$ in $k$-core percolation}
We consider $k$-core subgraph of a given ER network of size $N$ and mean degree $z$. If $z$ is larger than a critical value $z_c$, a $k$-core subgraph of size $M(z)N$ can exist. $M(z)$ 
was obtained analytically in \cite{kcore_goltsev}. We define $P_d(q)$ as the probability that a node in the $k$-core subgraph has degree $q$. We consider the evolution of the avalanche process after removing $\ro N$ nodes from the $k$-core subgraph. We define $u_{n}$ as the probability that a node attached to the end of an randomly chosen edge of a network will have degree less than $k$ at time step $n$. Then the evolution of $u_n$ satisfies the following equation,
\begin{equation}\label{key}
u_{n+1}=i_0+(1-i_0)
\sum_{q=k}^{\infty}\dfrac{qP_d(q)}{\langle q\rangle}f_q(u_n),
\end{equation} 
where $i_0 = \ro/M(z)$ and
\begin{equation}\label{key}
f_q(u_n)=\sum_{i=q-k+1}^{q-1}\binom{q-1}{i}{u_n}^i(1-u_n)^{q-1-i}.
\end{equation}
Taking similar steps as for the SWIR model, we now consider the avalanche process after a certain time step $m$, which is 
\begin{equation}\label{eq:kcoreEvo}
h_{n+1,m}=i_m+\sum_{\ell=k}^{\infty}Q_m(\ell) f_\ell(h_{n,m}),
\end{equation}
where
\begin{equation}\label{key}
h_{n,m}=\dfrac{u_n-u_{m-1}}{1-u_{m-1}},
\end{equation}
\begin{equation}\label{key}
Q_{m+1}(\ell)=\dfrac{1-i_0}{1-u_m}\sum_{q=\ell}^{\infty}\dfrac{qP_d(q)}{\langle q \rangle}
\binom{q-1}{q-\ell}{u_{m}}^{q-\ell}(1-u_{m})^{\ell-1},
\end{equation}
\begin{equation}\label{key}
i_{m}\equiv \sum_{\ell=1}^{k-1}Q_{m}(\ell)=\dfrac{u_{m}-u_{m-1}}{1-u_{m-1}}.
\end{equation}
Here $Q_{m}$ denotes the probability that a node attached to the end of randomly chosen edge in the remaining graph at time step $m$ has degree $\ell$.

We now consider sample to sample fluctuations at the characteristic time $n^*$. At a transition point $\ro=\rho_c$, Eq.~(\ref{eq:kcoreEvo}) has a nontrivial fixed point $u_d <1$. Defining $h_{d,n^*}\equiv (u_d-u_{n^*-1})/(1-u_{n^*-1})$ and $\epsilon_n=h_{n,n^*}-h_{d,n^*}$, Eq.~(\ref{eq:kcoreEvo}) with fluctuations becomes
\begin{eqnarray}\label{key}
h_{d,n^*}&&+\epsilon_{n+1} = \nonumber \\
&&i_m+\delta i_m+\sum_{\ell=k}^{\infty}\Big(Q_m(\ell)+\delta Q_m(\ell)\Big)
f_\ell(h_{d,n^*}+\epsilon_n) \nonumber \\
\end{eqnarray}
which can take the form of Eq.~(\ref{eq:e_n2}) with
\begin{equation}\label{key}
d_0=\delta i_{n^*}+\sum_{\ell=k}^{\infty}\delta Q_{n^*}(\ell)f_{\ell}(h_{d,n^*}).
\end{equation}

\end{document}